\documentstyle[prl,aps,epsfig]{revtex}

\begin{document}
\draft

\wideabs{
\title{Slow light and the phase of a Bose-Einstein condensate}
\author{P. \"Ohberg}
\address{School of Physics and Astronomy, University of St Andrews,}
\address{North Haugh, St Andrews, Fife, KY16 9SS, Scotland}
\maketitle
\begin{abstract}

We investigate the propagation of light with ultra low group velocity
in a Bose-Einstein condensate where the phase is not uniform. The light is shown
to couple
strongly to the phase gradient of the condensate. The interaction 
between the light and the condensate enables us to perform a phase imprinting
where the phase of the condensate is imprinted on the light. We illustrate the effect 
by showing how one can measure the fluctuating phase  
in an elongated quasicondensate.

\end{abstract}
\date{today}
}


Superfluids such as Bose-Einstein condensates of alkali atoms \cite{BEC} are
systems which show a conceptual simplicity yet provide all the intriguing
qualities of an interacting many-body system. Many superfluid phenomena are
strongly connected to the phase of the fluid. For instance
vortices \cite{Vortices} which have been created in a stirring manner similar to 
the rotating bucket experiment \cite{Tilley}, and solitons \cite{hannover,sol2},
are typical examples of a condensate with a nonuniform phase. Also quantum 
shock waves causing topological defects in the superfluid, have been created 
using ultra-compressed pulses of slow light \cite{Dutton}.
Here we show how the properties of slow light
can be used to detect the phase of the macroscopic wavefunction
of the condensate nondestructively, where the slow light receives the phase of the
condensate in a way similar to phase imprinting \cite{hannover,sol2}.


The recent experimental demonstrations with slow light has 
revealed a whole new playground with interesting applications
\cite{Dutton,hau,LP}. In this paper we show the advantages with slow 
light when the phase of the condensate is of interest. Light, and
especially the phase contrast imaging method, has been successfully used 
to probe the density of the condensate {\it in situ} \cite{Andrews}. In order to 
measure the phase, on the other hand, destructive methods have been 
used where the interference pattern has revealed the phase of the
condensate \cite{Inouye}. In this paper we show using slow light that the phase 
of the condensate can be imprinted on the light in a nondestructive
way, making it possible to follow the phase evolution of the condensate
{\it in situ}.

The envisioned experiment is based on Electromagnetically Induced
Transparency (EIT) \cite{EIT,leo,light}. Suppose that the condensate
is trapped in an elongated trap and illuminated by a uniform 
control beam. The beam controls the group velocity $v_g$ of a 
second beam, the probe beam. The probe shall be a pulse 
with a frequency in the laboratory frame that matches exactly the 
atomic transition frequency between the ground state and one of the
excited states coupled by the control beam. The condensate density,
which can be monitored {\it in-situ} by phase-contrast 
microscopy \cite{Andrews}, is supposed to be stationary, but the 
phase can in certain cases fluctuate \cite{stefanie,dima2}. For low group velocities the 
probe beam couples strongly to the condensate phase gradient 
which results in a phase imprinting on the light. The condensate
phase can consequently be read out by analyzing the interference pattern
between the initial and final probe light.

Let us start by deriving the equations of motion for the slow light and 
the Bose-Einstein condensate. The probe light is described by a real scalar field
$\varphi$ where we ignore the polarization, and $\varphi$ represents
the electric field in units of the vacuum noise,
$E=(\hbar/\varepsilon_0)^{1/2}\omega_0\varphi$. Here $\omega_0$
denotes the resonance frequency of EIT. In a medium at rest, slow
light is subject to the principle of least action with the
Lagrangian density \cite{leo}
\begin{equation}
{\cal L}_{L}=\frac{\hbar}{2}\left((1+\alpha)(\partial_t \varphi)^2-c^2(\nabla 
\varphi)^2-\alpha \omega_0^2 \varphi^2\right) \label{lag}
\end{equation}
where the group index $\alpha$ corresponds to a group velocity of
\begin{equation}
v_g=\frac{c}{1+\alpha}
\end{equation}
and $\alpha$ is proportional to the density of the condensate, $\rho$, 
and inversely proportional to the intensity of the
control beam \cite{Fleisch,leo},
\begin{equation}
(1+\alpha)\frac{\varepsilon_0|E_0|^2}{\hbar\omega_0}=\frac{1}{2}
\left|\frac{\Omega_p}{\Omega_c}\right|^2 \rho.
\end{equation}
Here the intensities of the probe and control
fields are calibrated in terms of the Rabi frequencies $\Omega_p$ and
$\Omega_c$, respectively. The control beam will dominate and in
practice $|\,\Omega_p |^2/|\Omega_c|^2$ does not exceed
$10^{-1}$ \cite{hau}. This means the less intense the control beam is 
the slower the light is.
The corresponding momentum density of the light
which is needed when describing the coupling between the light and 
the condensate, is obtained from the symmetric energy-momentum tensor
\cite{LL2,leo} 
\begin{equation}
{\bf P}=-\hbar (\partial_t \varphi)\nabla\varphi \,.
\end{equation}

Consider for the time being the general situation with a moving condensate.  
The condensate with 
the flow ${\bf u}$ and the probe light must now be taken as a combined dynamical
system with the total Lagrangian density
\begin{equation}
{\cal L}={\cal L}_L+{\cal L}_M \label{lagtot}
\end{equation}
where the condensate is described by the Gross-Pitaevskii Lagrangian density
\begin{equation}
{\cal L}_M=-\rho\left(\hbar \dot S+\frac{m}{2} u^2+\frac{\hbar^2}{2m}(\nabla \sqrt{
\rho})^2+\frac{g}{2} \rho+V\right).
\end{equation}
Here $m$ is the atomic mass, $g$ characterizes the atom-atom collisions, $V$ 
denotes the external potential and $\rho$ the condensate density. 
The coupling between light and matter is given by 
the relation \cite{ablight}
\begin{equation}
m {\bf u}=\hbar\nabla S+\alpha_0 {\bf P},\quad\alpha=\alpha_0 \rho \label{coup}
\end{equation}
where $\alpha_0$ is a constant. This coupling gives the correct equation of 
motion
\begin{equation}
\partial_t \rho+\nabla (\rho {\bf u})=0
\end{equation}
from the Euler-Lagrange equation. 

Using the Lagrangian in Eq. (\ref{lagtot}) we derive the equation of motion
for the light field. The resulting wave equation is
\begin{eqnarray}
\left( (1+\alpha)\partial_t^2-c^2\nabla^2+\alpha\omega_0^2+\alpha\partial_t {\bf u}\cdot
\nabla\right.&& \nonumber \\ \left. +\nabla\cdot\alpha{\bf u}\partial_t\right)
\varphi({\bf r})&=&0. 
\end{eqnarray}
In order to simplify things we assume the light field
can be expressed as a propagating pulse with a single frequency $\omega_0$ and a slowly varying 
amplitude. The resulting equation of motion is then of the Schr\"odinger type
\begin{eqnarray}
i\partial_t \varphi({\bf r})&=&-\frac{c^2}{2(\alpha+1) \omega_0}\nabla^2 
\varphi({\bf r})\nonumber\\&&
-\frac{i}{2}\frac{\alpha}{\alpha+1}
(2{\bf u}\cdot\nabla\varphi({\bf r})+(\nabla\cdot{\bf u})\varphi({\bf r}))
\label{ligeq}
\end{eqnarray}
where the coupling between light and matter is primarily described by the phase gradient 
of the condensate.

Let us illustrate the consequences of the coupling between the light and the phase of the
condensate by using an elongated quasicondensate where the phase can fluctuate 
\cite{stefanie,dima2}.
Fluctuations of the density and the phase of a condensate are related to
the elementary excitations. The density fluctuations are dominated by the 
excitations of the order of the chemical potential $\mu$. If the condensate
is very elongated but still of a 3D character, the wavelength of the density 
excitations are much smaller than the radial size of the condensate. The 
fluctuations are therefore of an ordinary 3D form and are small. Consequently
the total field operator can be written as
\begin{equation}
\hat\Psi({\bf r})=\sqrt{\rho_0({\bf r})}e^{i\hat\theta({\bf r})}
\end{equation}
where $\rho_0({\bf r})$ is the stationary density and
$\hat\theta({\bf r})$ is the operator describing the phase of the 
condensate. This operator is given by \cite{phase}
\begin{equation}
\hat\theta({\bf r})=\frac{1}{\sqrt{4\rho_0}}\sum_\nu f_\nu^+ \hat a_\nu +h.c.
\end{equation}
where $\hat a_\nu$ is the quasi particle annihilation operator with quantum
number $\nu$ and energy $\varepsilon_\nu$. The mode functions $f_\nu^+ =u_\nu+v_\nu$
are the sum of the two functions $u_\nu$ and $v_\nu$ which are the solutions 
to the Bogoliubov-deGennes equations.

In an elongated condensate the excitations are mainly of two kinds: axial 
excitations with $\varepsilon_\nu<\hbar\omega_\rho$ and radial excitations 
$\varepsilon_\nu>\hbar\omega_\rho$. The latter has a 3D character since the
wavelength is typically less than the radial size of the cloud and consequently
the fluctuations are small. The axial excitations on the other hand have wavelengths 
larger than the radial size of the cloud and have a 1D behavior. Therefore these 
excitations will be most important for the axial fluctuations of the phase. 
In order to actually calculate the phase we note that for a harmonic external potential
the low energy axial modes \cite{stringari} are described by the energy 
$\varepsilon_\nu = \frac{1}{2}\hbar \omega_z \sqrt{\nu (\nu+3)}$ and the functions
\begin{equation}
f_\nu^+ = \sqrt{\frac{(\nu+2)(2\nu+3)g\rho_0({\bf r})}{4\pi (\nu+1)R^2L\varepsilon_\nu}}
P_\nu^{(1,1)}(\frac{z}{L})
\end{equation}
where $P_\nu^{(1,1)}$ are the Jacobi polynomials, $R$ the radial size and $L$ the axial
length of the cloud. The quasiparticle annihilation operators are now replaced by 
complex amplitudes $\gamma$ and $\gamma^*$. 
To reproduce the quantum statistical properties of the phase the amplitudes are 
sampled as random variables with a zero mean value, 
$\langle\gamma_\nu\rangle=\langle\gamma_\nu^*\rangle=0$,
and bosonic number density 
\begin{equation}
\langle|\gamma_\nu|^2\rangle=\frac{1}{e^{\beta \varepsilon_\nu}-1}
\end{equation}
where $\beta$ is the inverse temperature \cite{stefanie}.

If the coupling is dominated by the phase gradient, in other words we 
neglect the nonlinear term for the light in Eq. (\ref{ligeq}), it is clear that we can 
use the light as a weak probe which is affected by the condensate phase gradient. The 
condensate will necessarily be very elongated. 
We can therefore solve the dynamics in 1D since the low energy axial excitations will acquire
a 1D character. From Eq. (\ref{ligeq}) it is 
immediately clear that the light will pick up a phase
\begin{equation}
\varphi(r,z)=\tilde\varphi(r,z)e^{i\xi(z)}
\end{equation}
where the the phase is given by the integral
\begin{equation}
\xi(z)=-\frac{k_0}{v_g} \int^z dx u(x)=-\frac{\hbar k_0}{m v_g} S(z)
\end{equation}
with $k_0=\frac{\omega_0}{c}$ and $S(z)$ the phase of the condensate. This 
scenario requires that the interaction between the probe light and the condensate
is turned on suddenly. This can indeed be achieved by tuning the control beam such
that the probe beam is rapidly slowed down and allowed to propagate in the condensate. 
This results in a pulse delay typically of the order of a few micro seconds \cite{hau}.

\begin{figure}
\centerline{\scalebox{0.6}{\epsffile{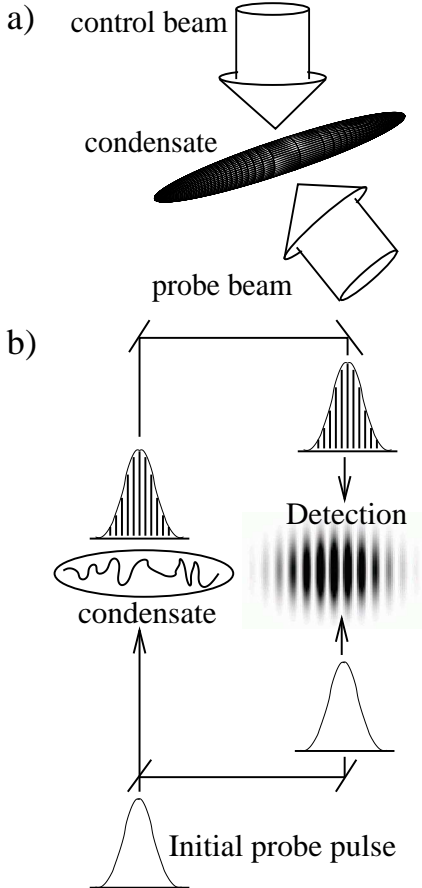}}}
\caption{a) The experimental setup with the control beam and the probe beam. 
The probe beam reveals the phase properties of the condensate.
b) The probe beam is split where one part is let to interact with the condensate.
Detection of the condensate phase is made with an interference experiment. Note 
the transversal interference pattern.}
\label{fig}
\end{figure}
\begin{figure}
\centerline{\scalebox{0.6}{\epsffile{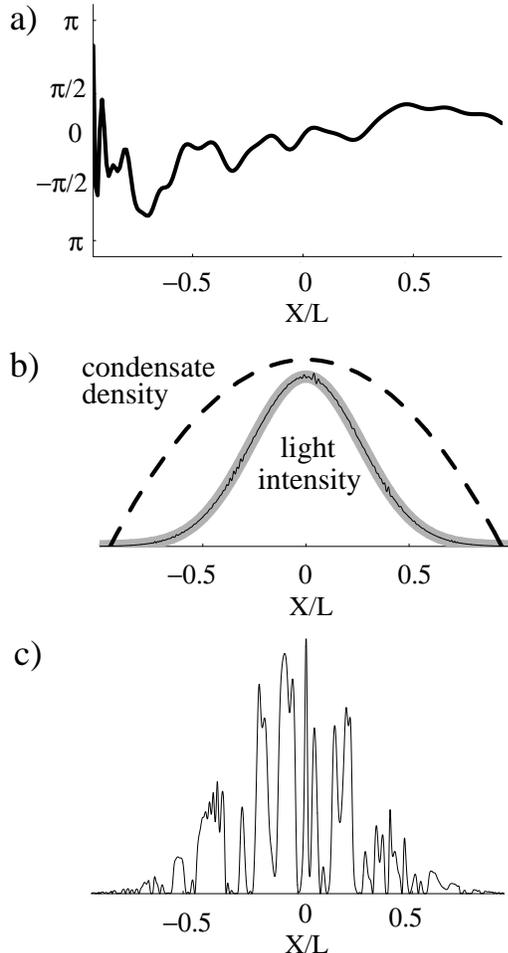}}}
\caption{(a) One possible phase at the temperature $T=0.75 T_{BEC}$. Only the lowest 
modes are important in the phase fluctuations. (b) The shape 
of the light pulse (solid black line) is only slightly altered due to the acquired phase. 
The grey curve shows the initial pulse shape (arb. units). The dashed line shows the density
of the condensate (arb. units) (c) The resulting intensity of the light when interfering 
with the initial probe light (arb. units).}
\label{res}
\end{figure}

To illustrate the phase measuring process we solve Eq. (\ref{ligeq}) for a light
pulse moving with the group velocity $1 mm/s$ during $1.5 ms$ in a condensate with 
$N=3.3 \times 10^4$ Na atoms, $\omega_z=2\pi\times 10 Hz$ and $\omega_r=2\pi\times 400 Hz$.
We choose a low group velocity and a relatively
long propagation time in order to show any effects on the light intensity from the 
condensate phase. In Fig.(\ref{res}) 
the condensate together with one possible phase representation is shown at a 
temperature well below the condensate critical temperature, $T=0.75T_{BEC}$. From
Fig. (2a) it is clear that only the low excitations are significant, as expected.
The phase
fluctuations are clearly present and are typically of the order of $\pi/3$ in the 
center of the condensate. The shape of the light pulse, in Fig. (2b), is only 
slightly altered due to the acquired
condensate phase. The pulse shape changes are typically small compared to
the overall pulse width and are therefore not necessarily very well suited for 
detecting any acquired phase from the condensate.  
In Fig. (2c) we show the resulting intensity when the light which has acquired the 
phase of the condensate, and the initial probe beam interferes. Note here that the
interference pattern will be in the transverse direction compared to the pulse propagation.
For sufficiently low group velocities the acquired condensate phase is amplified 
by the factor $\hbar k_0/m v_g$, which makes it possible to study also small
phase fluctuations not detectable by a time of flight measurement \cite{stefanie}.
As in any measurement process, there is inevitably going to be a back-action on the condensate
from the light. This back-action can, however, be made arbitrarily small by choosing 
an appropriate ratio $|\Omega_p|/|\Omega_c|$. The group velocity, which is a function
of the control beam, is not affected by this ratio. 

The 1D character of the condensate-light phase imprinting has here been used for 
simplicity. Similar phase fluctuations in 2D condensates can also exist \cite{dima} and 
consequently the same situation emerges concerning the properties of the light.
One important factor, however, is the inherent time dependence of the phase
fluctuations in the condensate. As was already seen in Fig. (\ref{res}) the 
phase fluctuations are dominated by the lowest modes. This means also that
the dynamics is governed by the corresponding frequencies 
$\omega_\nu=\varepsilon_\nu/\hbar$. For a typical external harmonic potential with
$\omega_z=2\pi\times 10 Hz$ we obtain a timescale for the phase fluctuations
of the order of $20 ms$, corresponding to the ten lowest modes, which is long 
enough to treat the phase as stationary in the imprinting process. The typical
timescale to apply and slow down the light pulse is of the order of micro seconds
\cite{hau} which allows us to assume an instant phase imprinting.

In summary we have shown that light with extremely low group velocity can be used 
to probe the phase of a condensate. The method was illustrated by 
imprinting the fluctuating phase of a quasicondensate onto the slow light propagating
through the condensate. The phase imprinting technique also allows for studying
other forms of phase gradients such as vortices \cite{paris} and solitons \cite{P+L}. 
Especially solitons in two-component condensates with repulsive interaction 
could be detected without opening the trap \cite{hannover}. Two-component solitons 
are significantly more difficult to observe because the total density will always 
stay constant, compared to the single component condensate soliton which shows a
density notch at the position of the soliton. The probe light in this case would 
simply pick up the phase of the soliton solution. If the light propagates
during a long time in a medium with a nonuniform phase it is also possible to study 
scattering of the light from the condensate phase \cite{ablight} which in principle can also
be used to measure the phase gradient, although this method will not necessarily 
reveal the condensate phase as clearly as a direct interference experiment. 

This work was supported by EPSRC. The author wishes to thank T. Kiss and 
U. Leonhardt for fruitful discussions. The Department of Physics at \AA bo Akademi
University is acknowledged for their hospitality.


\end{document}